\documentstyle[12pt]{article}
\begin{document}

\title{Observational Cosmology: caveats and open questions in the 
standard model}

\author{Mart\'\i n L\'opez-Corredoira\\
Astronomisches Institut der Universit\"at Basel\\
Venusstrasse 7. CH-4102-Binningen (Switzerland)}

\maketitle

{\bf \large ABSTRACT}

I will review some results of observational
cosmology which critically cast doubt upon the foundations of the 
standard cosmology: 
1) The redshifts of the galaxies are due to the expansion of the Universe;
2) The cosmic microwave background radiation and its anisotropies come from 
the high energy primordial Universe;
3) The abundance pattern of the light elements is to be explained in
terms of the primordial nucleosynthesis;
4) The formation and evolution of galaxies can only be explained in
terms of gravitation in the cold dark matter theory of an expanding Universe.
The review does not pretend to argue against this standard scenario in favour of an
alternative theory, but to claim that cosmology is still a very young science and should
leave the door wide open to other positions.

\

\section{INTRODUCTION}

Cosmology is a difficult science, if it can be considered a science
at all. It is young, it began to be considered as a science
less than a century ago. It is also very speculative, because
one has to reconstruct the complete history of the Universe with
few indirect observations. However, it is claimed as a science in all
aspects, and even a science with a final
explanation of the Universe in which only the fitting of few parameters remains
to be done. Disney's opinion\cite{Dis00} that cosmological inferences should be
tentatively made and skeptically received is something
which has been little respected in recent years by many 
cosmologists, who mostly believe that they have really this
final answer.

Discussions about possible wrong statements in the foundations of
standard interpretation (``Big Bang'' hypothesis) are normally not welcome. 
Many cosmologists are pretty sure that they have the correct theory and 
they do not need to think about possible major errors in the basis
of their standard theory. Most works in cosmology
are dedicated to refining small details of the standard model and do not
worry about the foundations.
From time to time, few papers are published which want to corroborate
that the foundations of their theory are correct through some cosmological
tests. They usually claim that a definitive proof is finally achieved. 
This is what I want to examine here.

In this paper, I will review critically the most important assumptions
of the standard cosmological scenarios\cite{Pee91,Nar92,Bar94,Nar01}: 

\begin{itemize}

\item The redshifts of the galaxies are due to the expansion of 
the Universe.

\item The cosmic microwave background radiation and its anisotropies come from
the high energy primordial Universe.

\item The abundance pattern of the light elements is to be explained in
terms of the primordial nucleosynthesis.

\item The formation and evolution of galaxies can only be explained in
terms of gravitation in the cold dark matter theory of an expanding Universe.

\end{itemize}

Some observations will be discussed or rediscussed in order to show that
these facts were not strictly proven. The possibility of
alternative hypotheses different from the Big Bang theory is still open. 
There are many alternative theories: Quasi-Steady State Theory\cite{Hoy93,Hoy94}, 
plasma cosmology\cite{Alf81,Alf83,Alf88}, Hawkins' model of eternal 
Universe\cite{Haw60,Haw62a,Haw62b,Haw62c}, chronometric cosmology\cite{Seg76,Seg95},
fractal Universe\cite{deV70,Pie77}, cold Bang\cite{Lay90}, 
wave system cosmology\cite{And99}, etc.
However, the mission of this review is not to analyze the different
theories, but mainly the observational facts. It is not my purpose to defend
a particular theory against the standard cosmology. All theories have
their own problems\cite{Pee93,Wri??,Nar01}, and will not be discussed here in detail.
Only the problems of the standard Big Bang theory are put forward.
Not all the problems and all the papers are considered since the 
literature is too vast in this wide topic.
I have chosen to do a review of the general aspects of the 
foundations in cosmology as a whole instead of some branch of it because I am interested in
expressing the caveats and open questions as a whole, to extract
global conclusions of cosmology as a whole. 
It may also be that some of the presented caveats are not caveats anymore, or
that some of the observational measurements are not correct. {\bf 
Warning:} {\it I just review some critical papers, and in some few cases
I discuss them, but I do not take responsibility for their contents.}
My own position is also neutral, I do not have any idea on whether 
the standard cosmology is correct or not. Maybe it is, maybe it is not, who knows? 
At present, I just want to express my opinion illustrated with the multiple
references of this review: we have not found the answer,
cosmology is a very young science.

\section{REDSHIFT, EXPANSION AND CONTROVERSIES}

\subsection{Does redshift mean expansion?}
\label{.redshift}

Hubble\cite{Hub29} and Humason\cite{Hum29} established in 1929 the 
redshift--apparent 
magnitude relation of the galaxies, which gave
observational proof that the universe is expanding.
Hubble was cautious in attaching significance to his own findings
about the law with his name and its implication of an expanding Universe.
It is indeed well known that Hubble
himself was disillusioned with the recession interpretation: {\it ``it seems 
likely that red-shifts may not be due to an expanding Universe, and much 
of the speculation on the structure of the universe may require 
re-examination''}\cite{Hub47}.
Perhaps Hubble was not so convinced by the idea of the expansion
of the Universe, but following generations decided to claim that
Hubble's discovery is a proof of the expansion, due mainly 
to the absence of a good theory which explains the possible phenomenological
fact of alternative proposals. General relativity provided an
explanation for the cosmological expansion, while alternative proposals were not supported
by any well-known orthodox theory. The expansion was preferred and the phenomenological
approaches which were not supported by present-day theory would
be doomed to be forgotten. This position would be right 
if our physics represented all the phenomena in the Universe, but
from a deductive-empiricist point of view we should deduce theories
from the observations, and not the opposite.

There are other mechanisms to produce redshift\cite{Nar89,Bar94}
beside space expansion or Doppler effect. There are many theories, for example: 
gravitational redshift\cite{Bon47,Bar81,Bar94b,Bar94}, chronometric
cosmology\cite{Seg76,Seg95}, variable mass hypothesis\cite{Hoy64,Nar77,Nar93},
inertial induction\cite{Gho97}, time ac\-ce\-le\-ra\-tion\cite{Gar01}, imperfect
photon propagation (A. Stolmar, priv. comm.), or the ``tired light'' scenarios.

A tired light scenario assumes that the photon loses energy due
to some unknown process of photon-matter or photon-photon interaction, 
when it travels some distance: long distance if we consider all the 
intergalactic space from the object to the earth; or short distance, 
for instance only the coronae which wraps
this object. Indeed, it is not a theory, it is a possible
phenomenological approach to explain the loss of the energy of the photons, which could be
explained with different theories. There are several hypothetical theories which
can produce this ``tired light'' effect. The idea of loss of energy of the photon
in the intergalactic medium was
first suggested in 1929 by Zwicky\cite{Zwi29} (see a review in \cite{Pec76}) and
was defended by him for a long time. Nernst in 1937 had developed a model
which assumed that radiation was being absorbed by luminifereous ether.
As late as the mid-twentieth century, Zwicky\cite{Zwi57} maintained that
the hypothesis of tired light was viable.
But there are two problems\cite{Nar89}: 1) the $\phi $-bath 
smears out the coherence of the radiation from the source, and so all
images of distant objects look blurred if the intergalactic space
produces scattering; 2) the scattering effect and
the consequent loss of energy is frequency dependent.
Vigier\cite{Vig88} proposed a mechanism in which the vacuum behaves like 
a stochastic covariant superfluid aether whose excitations can interfere
with the propagation of particles or light waves through it in a
dissipative way. This avoids the two former difficulties: the blurring
and the frequency dependence. The ``Incoherent Light Coherent
Raman Scattering''\cite{Mor01} also explains shifts which emulate
Doppler effect with light-matter interaction which does not
blur the images. The justification of the shift of photon frequency in a low
density plasma could also come from quantum effects derived from standard quantum
electrodynamics\cite{Lai97}. 
According to Paul Marmet and Grote Reber (a co-initiator of radio astronomy),
quantum mechanics indicates that a photon gives up a tiny amount of
energy as it collides with an electron, but its trajectory does not
change\cite{Ler91}(appendix). This mechanism also avoids
blurring and scattering. Potentially, this effect could explain the 
high redshifts of apparently nearby QSOs (Quasi Stellar Objects) 
(see \S \ref{.QSO}), since light traveling through the outer
atmosphere of the QSO could be redshifted before leaving it.
In order to explain galactic redshifts with long travel distances in the
scattering, the density in the intergalactic medium
should be 10$^4$ atoms/m$^3$, which is much higher than the density 
which is normally believed for it ($\sim 10^{-1}$ atoms/m$^3$). 
However, the disagreement in the density of the intergalactic
medium is not necessarily a caveat in the hypothesis since our
knowledge of the intergalactic medium is very poor, and there is
also the possibility that the intergalactic space is not so empty of
baryonic matter\cite{Lop99}. 

The dynamic multiple scattering theory is also very interesting
for the present question, as a possible tired light mechanism.
Results in statistical optics\cite{Wol86,Roy00} have shown that the shift
in frequency of spectral lines is produced when light
passes through a turbulent (or inhomogeneous) medium, due to multiple 
scattering effects. The new frequency of the line is proportional to the
old one, the redshift does not depend on the incident
frequency\cite{Roy00}. Several experiments have been succesfully
conducted in terrestrial laboratories leading to redshifts exceeding
300 Km/s \cite{Mey03}. The blurring which produces
this theory may be a problem when we take the whole intergalactic medium
as the substance which produces the shift but not if we consider some
loss of energy in the same coronae of the object. 

All these proposed mechanisms show us that it is quite possible to
construct a scenario with non-cosmological redshifts. 
Nonetheless, all these theories are at present just speculations 
without direct experimental or observational support.

\subsection{Observational tests on the expansion of the Universe}
\label{.testexp}

Since we cannot exclude alternative scenarios to explain the redshift of the
galaxies from a strictly theoretical
point of view, we must see whether there is a test or observational
evidence which is incompatible with a static Universe apart from the own
redshift of the galaxies.

Sandage\cite{San97} pointed out three different tests to verify whether the
Universe was really expanding:

\begin{enumerate}

\item Tolman surface brightness test.

Hubble and Tolman\cite{Hub35} proposed the so-called Tolman test based on the
measure of the brightness test. A galaxy at redshift $z$ differs
in the surface brightness depending on whether there is recession or not.
The different expressions are:
$\log I=$constant$-\log(1+z)-0.4K$ without expansion, and 
$\log I=$constant$-4\log(1+z)-0.4K$ with expansion,
where $K(z)$ is the K-correction.
This is independent of the wavelength.

Although simple in principle, the test is difficult to carry out because
galaxies as the test objects do not have uniform surface brightness.
The solution to this problem has been proposed in principle by using the so-called
Petrosian metric diameters with which to define an appropriate
area, and by finding the manner in which the average surface brightness
so measured varies with respect to the absolute magnitude of the E 
galaxies\cite{San90a,San90b,San91}.

In 1991, this test was carried out\cite{San91} and the results were not too clear. 
There was too much dispersion in their data to conclude anything. In spite of this,
Sandage declares in a congress about {\it ``Key problems in Astronomy''}\cite{San97} 
that the Tolman test would be feasible in few years.
Some authors\cite{Mol98} pointed out that this test is not appropriate 
because it cannot distinguish between the expansion and other effects which
give a dependence in the same direction but it was
replied by Lubin \& Sandage\cite{Lub01} that static and steady-state models 
were being confusing. 

Precisely Lubin \& Sandage\cite{Lub01} claim in 2001 to have a definitive 
proof of the expansion of the Universe using the Tolman test. Another definitive
proof in the long history of definitive proofs. However, I think that it
is questionable. Instead of finding that the surface brightness is 
proportional to $(1+z)^{-n}$ with
$n=4$, they found for a preferred cosmological model ($\Omega _M\approx 0.35$,
$\Omega _\Lambda \approx 0.65$) that $n=2.28\pm 0.17$ for the R band
and $n=3.06\pm 0.13$ for the I band. Since the values are different from
$n=4$ within several sigmas, the authors claim that the difference
is due to the evolution of galaxies. They speculate that the evolution
of galaxies gives precisely the factors [intensities brighter in the past
by factors: $(1+z)^{1.72\pm 0.17}$ for the
R band; $(1+z)^{0.94\pm 0.13}$ for the I band] which make compatible
their results with the expansion. Their claim is not 
a Tolman test but a claim that the evolution
of galaxies can explain the difference between the results of the Tolman
test and their preferred model which includes expansion.
Expansion model predicts a $(1+z)^{-4}$ dependence, 
and must therefore invoke special ad hoc evolution 
to close the gap between theory and observations.
This makes the Tolman test useless since we cannot separate its results
from the evolution of galaxies.

Lubin \& Sandage\cite{Lub01} claim to have {\it ``a definitive
proof that the hypothesis of non-expansion does not fit the
surface brightness data''}, in their own words. 
The Tolman test they carry out with the assumption of a static
Universe gives $n=1.61\pm 0.13$ for the R band and $n=2.27\pm 0.12$ for
the I band instead of $n=1$ which should be obtained for a static
Universe. Here, instead of saying that the difference is due to 
some other factor which explains the difference, they
claim that this is a definitive proof against a static Universe
because the evolution of galaxies cannot give negative luminosity evolution
in the look-back time, i.e. the evolution cannot explain why the galaxies
are fainter in the past. 
The authors have no problem in reconciling $n=2.28\pm 0.17$ with $n=4$ 
and cannot reconcile $n=1.61\pm 0.13$ with $n=1$ or $n=2$ 
(some tired light models might lose energy in the transverse direction, 
not just the longitudinal one, because light is a transverse wave;
when this effect is considered, the predicted loss of light intensity goes 
with $(1+z)^{-2}$). Even in the case of $n=1$,
I think that there is clear explanation for the galaxies to
be fainter in the past, and a static Universe is not excluded yet. 
First, I do not think that anything can be said about galactic
evolution, since the present models are derived from a Big-bang
scenario and that is precisely what we want to test. Second, I think
that the intergalactic absorption may be responsible for observing
fainter galaxies at farther distances. The intergalactic medium is there, and it 
has high abundance of metals\cite{Fan01,Pet01,Fan01b}, even at
high redshift (indeed no decrease of metallicity with redshift 
has been found so far\cite{Sch03}), so it is possible to have some dust 
and absorption too. We do not know how large it is but we must bear in mind that before we know it, 
we will not be able to contrast the different cosmological hypotheses. 
Errors due to inaccuracies of the method are also possible.

Therefore, I conclude that,
nowadays, there is not a definitive Tolman test which can be interpreted
directly without controversies in favour of one hypothesis or the opposite.

\item Time dilation test.

All clocks observed by us at large redshifts will appear to keep time at
a rate $(1+z)$ times slower when there is expansion. 
This has been known since it was proposed in 1939 by 
Wilson\cite{Wil39}. By using supernovae of type Ia, we would expect that 
their light curves would be stretched in the
time axis by a factor of $(1+z)$.
Until few years ago, no supernovae had been found at adequately large redshifts to
test the prediction.

The time dilation effect was recently measured\cite{Lei96}
on the type Ia supernova SN1995K with $z=0.479$ and the light curve was best 
fitted by a light curve stretched in the time axis by a factor of 
$(1+z)^b$, $b=1.0 ^{+0.50}_{-0.25}$. With a method to measure the age of the 
supernovae by analyzing their spectral features\cite{Rie97} a factor
of $(1+z)^b$, $b=2.4\pm 2.1$ was derived. A much better constraint to the value of
the parameter is the recent measure\cite{Gol01} of $b=1.07\pm 0.06$.
A possible criticism could be that 
the time under the light-curve depends on the intrinsic brightness of 
the supernovae, which might vary considerably with the redshift. 
However, it would be a queer coincidence that this effect produced $b\approx 1$
by chance, so I think the measures are a very interesting and solid
result about the detection of the time dilation.

These results seem to be in contradiction with the results
for QSOs\cite{Haw01}, which do not present time dilation as would
be expected if they were at the distance corresponding to the cosmological
redshift. If this analysis is correct, either the redshift of 
the QSOs is not cosmological (see \S \ref{.QSO}) or
these do not present the expected effect of the expansion.
With regard to GRBs (Gamma Ray Bursts) time dilation\cite{Cha01}, it is found that timescales tend to be 
shorter in bursts with small redshift, as expected from cosmological time 
dilation effects, although it is also found that there may be 
non-cosmological effects contributing to this correlation.

In any case, even if the time delay effect is present, as it seems to be
the case, this does not mean necessarily that the expansion is the only
explanation. Narlikar \& Arp\cite{Nar97} claim that the variable mass hypothesis
could also give a time dilation in supernovae light curves since the decay times
of the elements formed in these run on the slower clock times of lower
mass atoms, and they are dilated by exactly the $(1+z)$ factor. 
Segal\cite{Seg97} also explain the time dilation with his
chronometric cosmology.

Anyway, supernovae results are, in my opinion, the 
most impressive and solid test about the reality
of the expansion, and it might only be rejected if some alternative theory 
can emulate the time dilation too.

\item Microwave Background temperature as a function of the redshift.

Cosmic Microwave Background Radiation (CMBR, see \S \ref{.CMBR}) temperature 
can be detected indirectly
at high redshift if suitable absorption lines can be found in high redshift
objects. Indeed, Mackellar in 1941 had already measured the temperature of a
relic radiation of 2.3 K in our Galaxy to excite rotating cyan molecules before the Microwave
Background Radiation was discovered by other means\cite{Nov01}.
Hot Big Bang cosmology predicts that the temperature of CMBR is 
a function of the redshift. The prediction that the 
temperature required to excite these lines is
higher than in the Galaxy by the factor of $(1+z)$ can be tested, provided
that the radiation is in fact the one left over from the Big Bang.

By determining the excitation of atomic
transitions in absorbing clouds at $z=1.776$ along the line of sight to 
distant QSOs, assuming the transitions are in equilibrium with the CMBR, 
this temperature was calculated to be 7.4$\pm 0.8$ K \cite{Son94} which agrees with the
theoretical prediction of 7.58 K. However, another component of the cloud
with a very close redshift gave temperature 10.5$\pm 0.5$ K. 
Similar results are reported in \cite{Ge97}. Measures for a cloud at
$z=2.34$ give a temperature between 6 and 14 K \cite{Sri00}.

From the analysis of the C+ fine-structure population ratio in the damped 
Ly-$\alpha $ system at $z=3.025$ towards a QSO 
a temperature $14.6\pm 0.2$ K \cite{Mol02} is derived, while the theoretical
prediction is T=10.97 K. Since the agreement was not achieved directly,
they attribute the difference to the existence of other mechanisms
of excitation, specially the collisional. This means that the previous
measures in other papers should also be affected by other mechanisms of
excitation and they can just give the maximum CMBR temperature, but
not the minimum. Therefore, previous works have not proven the increase 
of CMBR temperature with $z$ by a factor $(1+z)$.
According to Molaro et al.\cite{Mol02}, UV pumping or IR dust emission are not an important 
excitation mechanism for C+, so they neglect such a contribution. 
When they consider collisional excitation ($n_H$ between 4 and 14 cm$^{-3}$), 
a temperature for the cosmic background radiation 
of $T_{CMBR}=12.1^{+1.7}_{-3.2}$ (95\% C.L.) K is measured. 
However, this is an indirect measure which depends on the model and
the calculations of the collisional excitation, which have some uncertainties.
These detections could also be interpreted as another radiation 
not associated with the CMBR. They allow to determine the maximum temperature
of the CMBR but the minimum temperature depends on the corrections
due to collisional excitations or others, which depend on many other parameters.
The result depends on the estimate of other contributions to the excitation,
so nothing can yet be concluded definitively from this test.

\end{enumerate}

Apart from these tests proposed by Sandage\cite{San97}, there are some other
tests in the literature:

\begin{itemize}

\item Apparent magnitude vs. distance test.
The Hubble diagram for elliptical galaxies in clusters\cite{LaV86,Sch97}, 
whose absolute magnitude and radii are related by the fundamental plane, taking
into account K-corrections, do not fit the expanding 
Universe for $q_0=1/2$ or $q_0=0$ but the static Universe.
The present values of $\Omega _m\approx 0.3$ and $\Omega _\Lambda \approx 0.7$
change the values of $q_0$ somewhat, but
still it is not enough to fit the data. 
This disagreement could however be solved by
an increase of luminosity at higher redshift due to the evolution
of galaxies (which is a possible observational result
pointed out from deep infrared FIRES and 
HST data\cite{Tru03}), so the uncertainties in the luminosity evolution 
do not allow this to be a proof against expansion.

\item Angular size vs. redshift test. 
Tests were made in the last 25 years by 
several groups\cite{Kap77,Djo81,LaV86,Kap87,Kel93} 
with different values of $q_0$. Tests were also made with
more recent HST observations of first-rank elliptical galaxies\cite{Sch97,Pas96}.
The angular size of radio galaxies over a range up to redshift 2
shows a dependence $\theta \sim z^{-1}$ \cite{Kap87,And99}, 
a static euclidean effect over all scales. 
This result cannot be reconciled with 
Cosmological Hypothesis unless we assume galactic evolution which changes
the radius of the galaxies for different redshift with an appropriate
dependence. However, the fit of radio source counts 
with no evolution is best\cite{Das88}.
The critical assumption of a variable effective
radius is also counter-argumented by the proofs in favour of a constant
radius\cite{van98} showing that high redshift first-rank
elliptical galaxies have the same velocity dispersions as low redshift 
first-rank elliptical galaxies. Because the velocity dispersion scales
with the mass, all first-rank elliptical galaxies must have nearly the same
mass. Then, since size is correlated with mass, first-rank elliptical
galaxies must have approximately the same effective radius.
The similar absolute magnitudes of the first-rank elliptical galaxies
at high $z$ and low $z$ also confirms that all first-rank elliptical galaxies
have similar effective radii (T. B. Andrews, priv. comm.).
Therefore, this test seems to show evidence more in favour of a static
Universe than one in expansion.
However, the main problem of such a type of test is the identification of the
``standard stick'' for which one knows reliably its non-variation with
redshift, or its variation with a well-known evolutionary scheme
so we must be cautious too to interpret the results of this test as
a definitive proof of a particular cosmological model, in this case
the static Universe. And, the variation of the size of the galaxies
with redshift, in spite of the above given arguments, might be a fact (as it is
pointed from FIRES-HST data\cite{Tru03}), so again these
tests are not definitive.

\item Temperature of the intergalactic medium.
The falloff of the power spectrum of the Ly$\alpha $ forest
at small scales can be used to determine the 
temperature of the intergalactic medium. It is typically inferred to be 
20000 K, but there is no evidence of evolution with redshift\cite{Zal01}. 
Yet in the Big Bang, that temperature ought to adiabatically decrease as space 
expands everywhere. If these measurements are confirmed by other means, it 
could be another indicator that the universe is not really 
expanding unless other explanations arise. At present it is early
to establish definitively this claim.

\end{itemize}

\subsection{Anomalous redshifts}

Doubt might be cast upon the reality of the expansion as it was discussed,
but even in the case that it is definitively proven, this does not mean that
all galaxies have a cosmological redshift, i.e. that all galaxies have a
redshift due to the expansion. There might be some exceptions, which are known
as ``anomalous redshift'' cases. In this subsection, I review some of
the papers which touch this question.

There are some cases in which ``apparently'' connected or interacting
galaxies have different redshifts\cite{Arp87,Nar89,Arp98}. 
One of the most famous cases is Stephan's quintet(SQ)\cite{Bur61}.
The system has one member (NGC 7320)
with redshift 0.0027 and the other four with redshifts 0.019, 0.022, 0.022,
0.022. Some authors\cite{Arp73,Bur61,Sul83,Arp87} 
have argued strongly against NGC 7320 as a foreground object with no physical proximity
to the other members of SQ. Furthermore, it is suspected that
the SQ along with a few other redshift companions are all
satellite members of NGC 7331, a large Sb galaxy with redshift nearly the same as
that of NGC 7320, because there is radio-emitting
material connecting NGC 7331 to the quintet\cite{Arp73}. However, an optical counterpart of this radio-emitting
material was not found up to magnitude 26.7 mag/arcsec$^2$\cite{Gut02}.
It has also  been claimed\cite{Mol01} that HST-images have definitely solved the
question of the redshift discrepancy of NGC 7320 because it is possible
to resolve individual stars in this galaxy, showing that it is closer
than the rest of the galaxies of the group. Arp\cite{Arp01}, 
in strong disagreement with this statement, claims that in these
images it is also possible to resolve individual stars in the
high-redshift members of the group. Furthermore, using the luminosity
of the H~{\scshape ii} region as a distance indicator, Arp\cite{Arp01} also
points out that the low- and high-redshift galaxies of
SQ should be at the same distance. There is also 
a H$_\alpha $-filament which is extending throughout NGC 7320 with velocity
at 6500 km/s instead of 800 km/s expected for this galaxy\cite{Gut02}, which is 
somewhat unlikely to be there by chance. The conclusions about SQ are not clear yet.

There are many other systems like SQ.
Hickson\cite{Hic82} has made a survey and has published a catalogue of 100
compact groups of galaxies containing four to six members. 
All the redshifts have been measured. Twenty-eight of the systems have 
one redshift very different from the mean of the others\cite{Bur97}. 
Although the numbers, sizes, magnitudes and morphological
types of the discordant redshift members may agree with a scenario
of chance projections, the distribution of
positions in quintets is more centrally concentrated than expected in
such a scenario\cite{Men95}.

There are also some photographic evidences
of some filamentary structure joining two galaxies: filaments which start
at one galaxy and end at the latter, suggesting that the compact member
was somehow ejected from the bigger object\cite{Arp87}.
Example: the Seyfert 1 galaxy NGC 7603 ($z=0.029$)
appears connected to a companion galaxy ($z=0.056$) and along the thin 
filament there are two further HII-galaxies with
redshifts 0.24 and 0.39\cite{Lop02,Lop03}, 
which makes the system much more unprobable in terms of cosmological redshifts.
The calculation of the probability by chance of this configuration gives numbers
of the order $10^{-9}$. Another interesting case is NEQ3\cite{Arp77,Gut03}:
a galaxy with $z$=0.12, a filament along its minor axis which ends in a group
of three objects with separations less than 3 arcsecond with respect the central
one: a QSO ($z$=0.19), a HII-galaxy ($z$=0.19) and a HII-galaxy ($z$=0.22).

\subsubsection{Apparent associations of QSO-galaxy with different redshifts}
\label{.QSO}

The number of these associations is even wider, although still
not widely accepted as such. 
Non-orthodox hypothesis shared by very few astronomers (Arp, Burbidge \& Burbidge
and few others) claim that QSOs are being ejected by low redshift active galaxies. 
Galaxies beget galaxies, they are
not made from initial density fluctuations in a Big Bang Universe\cite{Bur97b}. 
The redshift is intrinsic and quantized. Arp's model\cite{Arp99} assumes that QSOs
are ejected along the minor axis of the parent
galaxy decreasing in redshift as they move outward and becoming normal
galaxies when they are older. 
The narrow line character in some objects with anomalous redshift, like
the HII-galaxies cited in the previous subsection\cite{Lop03}, might be a result 
of QSOs which interact with gas of the filaments, or a different stage
of the same object in its evolution.
Some variations were also proposed: Driessen\cite{Dri96}, for example, 
claimed that QSOs are the particles exchanged in a new interaction among galaxies, 
something similar to the role of photon in the electromagnetic interaction.
Bell\cite{Bel02,Bel02d} thinks that each ejection event is in triplets, 
and inmediatly after the ejection each triplet splits into a singlet and a pair.
The main idea in any case is that QSOs are really 
at the same distance than nearby galaxies in spite of their high redshifts.

Evidence for the ejection origin of QSOs was first seen in the pairs of
radio sources across active galaxies. Radio sources were accepted as
ejected and a number of them were early identified as QSOs\cite{Arp67,Arp68}.
Statistical evidence continued to grow for such associations, and when
X--ray QSOs started to be discovered further examples
of possible associations with active galaxies were found\cite{Arp95}.
Strong statistical evidence for the association of
QSOs was found, at all redshifts, with galaxies with $z\le 0.05$\cite{Chu84},
the correlation being much higher for bright QSOs\cite{Gaz03}
(indeed, QSOs with small angular separation from a galaxy have in average
higher luminosities\cite{Dra96}). And there are many further examples of correlations 
of QSOs-galaxies or QSOs-QSOs with very different redshifts
with other catalogues\cite{Bur97,Bur97c}, some of them at the 7-10$\sigma $ 
level. Some authors (e.g., \cite{Slu03}, hypotheses H2-H3) 
get higher probabilities because they use in their 
calculation the limiting magnitude of the survey 
instead of the magnitude of the object. But this is not correct 
because randomly one would expect that most of the detected objects
were near the limiting magnitude. The magnitude of the correlated object 
tends to be much brighter than the limiting magnitude\cite{Dra96}, so 
one should multiply the probability by a factor which characterizes the 
probability to have this object much brighter than the limiting magnitude 
(the brighter it is, the lower is the probability), and this is equivalent 
to using the magnitude of the object.

All non-elliptical galaxies brighter than 12.8
with apparent companion galaxies were analyzed, and 13 of the 34 candidate 
companion galaxies were found to have QSOs of higher redshift
with an accidental probability of less than 0.01 for each one\cite{Arp81}; 
the global probability of this being a chance is $10^{-17}$.
Only one system like NGC 3842, with 3 QSOs at distance less than 73'',
have a probability lower than $10^{-7}$ to be a multiple projection
effect\cite{Arp84,Arp87}.
There are also configurations of QSOs aligned along the minor axis
of a central Seyfert whose probability is $10^{-9}$ to $10^{-10}$ of
being accidental\cite{Arp98b,Arp99}. And in some cases, the QSO is unlikely
very close to the galaxy to be a fortuitous projection: e.g.,
a QSO of redshift $z=1.695$ within 0.3
arcseconds of the nucleus of the galaxy 2237+0305 with $z=0.0394$ \cite{Huc85}.

There is anisotropy in the radio QSO distribution at high flux
densities\cite{Sha83}. The number of
QSOs in one side of the M33 region is far larger ($\sim 11\sigma $) than
that of the diametrically opposite region. 
The strongest concentration of QSOs with $z\sim 1$ is in an area of the sky
covering a solid angle of diameter 40 degrees 
apparently located in the Local Supercluster\cite{Nar89}.
Also, a grouping of 11 QSOs close to NGC 1068 (a Seyfert galaxy which
has itself very peculiar kinematics\cite{Cec02}: knots with blueshifted
radial velocities up to 3200 km/s, and gradients in radial velocities up to
2000 km/s in 7 pc) have nominal ejection patterns correlated with galaxy rotation,
the mean redshifts of the pairs fall off approximately linearly with increasing
distance from the Seyfert galaxy and are quantized\cite{Bel02,Bel02c}.

Detailed investigations of several QSO-galaxy pairs show that the
galaxy and QSO are apparently connected. For example, the connection by a
luminous bridge between NGC 4319 and Mk205\cite{Sul87,Mey03}.
It was argued that the filament is a photographic artifact, but
the system has been subjected to CCD pictures 
which has demonstrated the reality of the filament. 
Another case is the connection with radio contours of radio sources with $z$=0.34 
and $z$=0.75 \cite{Fan85}.
Or the neutral hydrogen clouds connection between the QSO 3C232 ($z$=0.5303) and 
the galaxy NGC 3067 ($z=0.0049$) \cite{Car89,Car92}.
I find this last case very amazing since the bridge in the radio contours which 
connect both objects is nearly 2 arcminutes long, with redshift associated to the 
NGC 3067 galaxy but with 
maximum intensity in the place where the QSO is placed.
NGC 3628 seems to have some relation with some surrounding QSOs\cite{Arp02}:
several QSOs are aligned across the nucleus of NGC 3628 
($z$=0.981 in one side, and $z$=0.995 and 
$z$=2.15 in the other side)---these are connected by the
X-ray contours with the nucleus of the galaxy---, they are within the
two plumes in HI belonging to the galaxy, and moreover there is an optical
filament connected with the galaxy which points in the direction towards the
QSO at z=0.995. Other cases of X-ray, radio and optical connections 
in \cite{Arp01b}. Nonetheless, there might be some of these cases 
in which the apparent bridge joining two structures is an artifact, 
either because the signal/noise of the contours is not significant\cite{Har00}
(e.g., some X-ray contours), or the superposition of contours of different objects 
can create these small bridges when the outer contours of both sources are very close 
(e.g., NGC 4319) or the survey space resolution is low, but
there are many of them (e.g., NGC 3067) which cannot be solved by any 
artifacts like these.

Weak gravitational lensing by dark matter was proposed to be the cause of the
correlations although this is very insufficient to explain them\cite{Ben01,Gaz03,Lop03},
and cannot work at all for the correlations with the brightest and
nearest galaxies\cite{Kov89}, even less to explain the bridges/filaments. 
As a matter of fact, the observed amplitude of the correlation is 
around 100 times the expected value for a weak lensing from the standard 
cosmological values\cite{Gaz03}.

\subsubsection{Other open questions about QSOs}

Other anomalies or questions with not very clear answers have arisen about
the nature of QSO and the suspicion that their distance is not as high as pointed out 
by the cosmological interpretation of their redshift.

The luminosity required for QSOs to be at such large distances is
between $10^{43}$ to $10^{47}$ erg/s, an enormous energy to be produced
in a relatively compact region (to justify the strong variability in short times). 
Although this problem is solved by means
of megahuge black-holes, the explanation might be related with a bad
determination of the distance too.
It seems that there are special problems to justify the abundance of very high luminosity
QSOs at $z\sim 6$, and the gravitational lensing solution does not work\cite{Yam03,Fan01b}.
Moreover, the huge dispersion in the magnitude-redshift relation for QSOs observed
from the Hewitt \& Burbidge catalogue\cite{Hew87} makes impossible to derive a
Hubble law for them. This is a not strong argument since the intrinsic dispersion
of luminosities might be high itself, but possibly it is pointing out that
something is wrong with the distance measure. 

Another caveat is that
the average luminosity of QSOs must decrease with time in just the right way 
so that their average apparent brightness is the same at all redshifts, which is
exceedingly unlikely.
On average, QSOs with very different redshift have comparable 
apparent brightness. This must be explained in the standard scenario
as QSOs evolving their 
intrinsic properties so that they get smaller and fainter as the universe 
evolves. That way, the QSO at $z=1$ must be intrinsically many times 
brighter than the one at $z=0.1$ to compensate its much higher distance, 
explaining why they appear on average to be 
comparably bright. Why QSOs should 
evolve in just this way to produce this coincidence of apparent magnitudes whatever
it is its redshift?\cite{van92}.
Furthermore, the number density of optical QSOs peaks at $z$=2.5-3, and declines toward 
both lower and higher redshifts. At $z=5$, it has dropped by a factor of 
about 20 \cite{Fan01b}. At $z=6$ is half of that at $z=5$ QSOs\cite{Fan01a}. 
The standard model predicts that QSOs, the seeds of all 
galaxies, should be more numerous at earliest epochs, so apparently there
is a problem here, which might be solved ad hoc, 
but not as a natural prediction of the model. 
The role of gas accretion for QSO activity might have something to do.
By contrast, the relation between apparent magnitude and distance for QSOs 
is a simple, inverse-square law with intrinsic redshifts.

Superluminal motions of distant sources ($D$) 
are observed, i.e. angular speeds $\omega $ between 
two radio emitting blobs which imply linear velocities $v=D\omega $ greater
than the speed light\cite{Coh86}. There are some explanations.
The so called relativistic beaming model\cite{Ree67} assumes that there is one blob
$A$ which is fixed while blob $B$ is traveling almost directly towards
the observer with speed $V<c$ with an angle $\cos ^{-1}(V/c)$ between
the line of advance and the line $B$-observer. This leads to an apparent
velocity of separation which may be greater than $c$. There is also another proposal 
in a gravitational bending scenario\cite{Chi79}. However,
both explanations share the common criticism of being contrived and
having somewhat low probability ($\sim 10^{-4}$)\cite{Nar84}.

Another problem with QSOs: PG 0052+251 is at the core 
of a normal spiral galaxy. The host galaxy appears undisturbed 
by the QSO radiation, which, in the standard scenario, is supposed to 
be strong enough to ionize the intergalactic medium\cite{Bah96}.

As said, QSOs do not present time dilation\cite{Haw01} as would
be expected if they were at the distance corresponding to the cosmological
redshift (if the expansion is real).

The polarization of radio emission rotates as it passes through magnetized 
extragalactic plasmas. Such Faraday rotations in QSOs should increase 
(on average) with distance. If redshift indicates distance, then rotation 
and redshift should increase together. However, the mean Faraday rotation 
is less near $z=2$ than near $z=1$ (where QSOs are apparently intrinsically 
brighter, according to Arp's model\cite{Arp98}).

The metallicity and dust content of very high redshift QSOs and their host galaxies 
is equal or even larger in some cases than their metallicity at low redshifts
(see \S \ref{.distgal}). 

The absorption lines of damped Ly$\alpha $ systems are seen in QSOs.
The HST NICMOS spectrograph has searched to see these objects
directly in the infrared, but failed for the most part to detect them\cite{Col02}. 
Moreover, the relative abundances have surprising uniformity, 
unexplained in the standard model\cite{Pro02} and, as said (see \S
\ref{.testexp}), the temperature of these clouds does not change with
redshift\cite{Zal01}, a fact which does not fit the normal
predictions of the standard model. 

It is difficult to obtain an unique conclusion with all this information.
At least, I think, they should serve to keep the doors open and not reject
automatically the idea that QSOs and possibly other galaxies might have
a non-cosmological redshift.

\section{MICROWAVE BACKGROUND RADIATION}
\label{.CMBR}

Gamow\cite{Gam53} and Alpher \& Herman\cite{Alp49} predicted an early stage of
the Universe in the Big Bang model which would produce a relic radiation
from this fireball that could be observed as a background.
However, Gamow and his coworkers were of the opinion that the
detection was completely unfeasible\cite{Nov01}.
The first published recognition of the relic radiation as a
detectable microwave phenomenon appeared in 1964\cite{Dor64}. 
None of the predictions of the background temperature based on the 
Big Bang, which ranged between 5 K and 50 K, 
matched observations\cite{van93}, the worst being 
Gamow's upward-revised estimate of 50 K made in 1961. 
In 1965, the year of the discovery, a temperature of 30 K was calculated
for the amount of helium production observed\cite{Dic65}.
This background radiation was discovered this year by Penzias \& Wilson\cite{Pen65}
with a temperature of 3.5 K and is nowadays measured as 2.728 K\cite{Mat94}.
As the energy is proportional to $T^4$,
the energy observed is several thousand times less than predicted energy,
but it was predicted correctly that
it has a perfect black-body spectrum\cite{Mat94}.
Although the discovery is attributed to Penzias \& Wilson, who won
the nobel prize because of it, the radiation was indeed previously 
discovered, although not interpreted in terms of a cosmological radiation:
in 1957, Shmaonov\cite{Shm57} was measuring radio waves coming from space at a wavelength
of 3.2 cm and obtained the conclusion that the absolute effective temperature
of radiation background appears to be 4$\pm $3 K, regardless of the 
direction of the sky. It is also possible that a team of japanese
radioastronomers measured this radiation at the beginning of 1950s\cite{Nov01}, and indirectly
it was found by MacKellar in 1941 as the necessary radiation to excite rotating
cyan molecules\cite{Nov01}. 
Anyway, the radiation is there and the question is the
origin of this radiation not the discoverer. 
Is the high energy primordial Universe the only possible scenario? 

Charles-Edouard Guillaume (nobel laureate in physics 1920) 
predicted in his article entitled ``Les Rayons X'' (X-rays, 1896) 
that the radiation of stars alone would maintain a background temperature 
of 6.1 K\cite{Mey03}. The expression ``the temperature of space'' is the title of 
chapter 13 of ``Internal constitution of the stars''\cite{Edd26,Mey03} 
by Eddington. He calculated the minimum temperature any body in space would cool 
to, given that it is immersed in the radiation of distant starlight. 
With no adjustable parameters, he obtained 3 K, essentially the same 
as the observed background (CMBR) temperature.
Other early predictions\cite{Mey03}, given by Regener\cite{Reg33} in 1933 
or Nernst\cite{Ner37} in 1937, gave a temperature of 2.8 K 
for a black body which absorbed the energy of the cosmic rays arriving on earth.
It was argued that Eddington's argument for the 
``temperature of space'' applies at most to our Galaxy. But Eddington's reasoning 
applies also to the temperature of intergalactic space, for which a minimum 
is set by the radiation of galaxy and QSO light. The original calculations 
half-a-century ago showed this limit probably fell in the range 
1-6 K\cite{Fin54}. And that was before QSOs were discovered 
and before we knew the modern space density of galaxies.
In this way, the existence of a microwave background in a tired light scenario\cite{Fin54,Bor54} was also deduced.
But in a tired light model in a static universe the photons 
suffer a redshift that is proportional to the
distance traveled, and in the absence of absorption
or emission the photon number density remains constant, we would not see a
blackbody background. The universe cannot
have an optical depth large enough to preserve a thermal background spectrum in
a tired light model\cite{Pee98} because we could not observe radio galaxies
at $z\sim 3$ with the necessary optical depth. Therefore, it seems that this
solution does not work, at least when the intergalactic medium instead of the
own shell of the galaxy is responsible for the tired light.

In the fifties, it was pointed out\cite{Bon55,Bur58} that if the observed
abundance of He was obtained by hydrogen burning in stars, there must have
been a phase in the history of the Universe when the radiation density
was much higher than the energy density of starlight today. If the
average density of the visible matter in the Universe is $\rho \sim
3\times 10^{-31}$ g/cm$^3$ and the observed He/H ratio by mass in it is 0.244 
(see \S \ref{.nucleo}), then the energy which must have been released in producing
He is $4.39\times 10^{-13}$ erg/cm$^3$. If this energy is thermalized, the black
body temperature turns out to be $T=2.76$ K, very close to the observed
temperature for the CMBR. Hence, there is a likely explanation
of the energy of the microwave radiation in terms of straightforward astrophysics
involving hydrogen burning in stars\cite{Bur97b}.
Hoyle et al.\cite{Hoy68} also pointed out the suspect coincidence
between the microwave background temperature and that of hydrogen
in condensation on grains. They postulate that galaxy and star formation
proceed very readily when hydrogen is condensed on grains but not when
it is gaseous, and the universal microwave background temperature is that
associated with galaxy formation. 
The mechanism of thermalization in any of these cases
is perhaps the hardest problem.
Dust emission differs substantially from that of a pure blackbody.
Moreover, dust grains cannot be the source of the blackbody microwave radiation 
because there are not enough of them to be opaque, as needed to produce 
a blackbody spectrum. The solution for the black-body emission shape
might be the special properties of the particles, which are not
normal dust particles: carbono needles, multiple explosions or big
bangs, energy release of massive stars during the formation of galaxies,...
The quasi-steady state model\cite{Hoy93,Hoy94} argues that there is a distribution of
whiskers with size around 1 mm long and 10$^{-6}$ cm in diameter
with average $\sim 10^{-35}$ g/cm$^3$ providing $\tau \sim 7$ at
redshift $\sim 4$. 
However, the presence of a huge dust density to make the Universe opaque 
is forbidden by the observed transparency up to $z\sim 4$ or 5.
Somewhat similar is the proposal of the thermalization of ``cosmoids'',
cosmic meteoroids which are also observed in the solar system\cite{Sob01}.
A solution might be an infinite universe.
Opaqueness is required only in a finite 
universe, an infinite universe can achieve thermodynamic equilibrium 
even if transparent out to very large distances because 
the thermal mixing can occur on a much smaller scale than quantum particles,
e.g., in the light-carrying medium itself. 

Another possible explanation is given by Clube\cite{Clu80} who considers the
existence of an aether, i.e. a material vacuum, whose emission gives the
microwave background emission itself. This aether would be an incompressible
fluid according to Lorentz's theory and its existence is 
in opposition to Einstein's special relativity. 
Lorentz\cite{Lor04} invariant laws of mechanics
in a flat space-time reproduce the standard observational tests of general
relativity provided that rest mass and light speed are not constants
so the final choice between Einstein and Lorentz theories cannot
yet be regarded as settled according to Clube\cite{Clu80}.

Lerner\cite{Ler88,Ler95} proposes that electrons in intergalactic magnetic fields
emit and absorb microwave radiation. There is no relation between the
direction in which the radiation was moving when it was absorbed and its direction
as a reemission, so the microwaves would be scattered. After a few scatterings,
the radiation would be smoothed out. Magnetic fields much
stronger than the average field between galaxies would be needed;
perhaps the jets emitted from galactic nuclei would provide it.
The background radiation would be 
distorted by this intergalactic absorption against isotropy observations, so 
the radiation must instead come from the intergalactic medium itself in
equilibrium. This prediction agrees with the fact that the number of
radio sources increases much more slowly than the number of optical sources 
with distance; presumably due to this absorption of radio waves in the 
intergalactic medium.
Observational evidence was also presented that something in the intergalactic medium
is absorbing radio and microwaves because farther radio sources with a given constant
infrared emission are fainter in radio\cite{Ler90,Ler93}.  
However, there are some sources which are quite
bright in radio at intermediate redshift: 
Cygnus A ($z=0.056$) or Abell 2218 ($z=0.174$).
There is even a constant FIR/Radio emission up to $z=1.5$ \cite{Gar02} and sources 
are observed at $z=4.4$, so unless we have
a problem of anomalous redshift in all these cases, which seems unlikely,
Lerner's ideas do not work.

At present there is not a satisfactory alternative scenario
which has no problem to explain the Microwave Background Radiation, so
the standard scenario seems the best solution. Nonetheless, neither the
standard scenario is free of caveats to explain all the facts. For example,
many particles are seen with energies over $60\times 10^{18}$ eV. 
But that is the theoretical energy limit for anything traveling for
more than 20-50 Mpc because of interaction with microwave 
background photons\cite{Sei00}. This is
a problem for the actual interpretation of CMBR in the Big Bang theory.
Therefore, one should not be definitively closed against other alternatives.
The hot primordial Big Bang is only a hypothesis which cannot be
definitively established as a solid theory due only to the existence
of this Microwave Background Radiation.

\subsection{Microwave Background Radiation anisotropies}
\label{.anis}

The Microwave Background Radiation has anisotropies, small variations of the
flux with the direction of the sky. First predictions of the anisotropies in Microwave
Background Radiation have given values of one part in hundred or thousand\cite{Sac67}; 
however, this could not fit the observations which gave
values hundreds times smaller, so non-baryonic matter 
was introduced ad hoc to solve the question. The same can be said of the
predictions of the peaks in the power spectrum of these anisotropies.
The acoustic peaks on angular scales of 1 deg. and 0.3 deg. were predicted
with the second peak nearly as high as the first one\cite{Bon87}. Successively,
as the data came, its amplitude was reduced.  
Newly acquired data from the Boomerang balloon-borne instruments\cite{deB00} shows
a very small or negligible signal of the second peak, so again the cosmological
parameters were refit. Apparently, the history
of Microwave Background Radiation is a set of predictions which do not
fit the observations but allow to modify the model to fit them ad hoc.
This is called ``successful predictions of the standard model''.

And again, like in other cases, we have a set of successful predictions
of other alternative theories, most of them also ad hoc.
Narlikar et al.\cite{Nar03} explains the observed anisotropies with the Quasi
Steady State Theory. The first peak in the power spectrum at $l\approx 200 $ 
(considered the first acoustic Doppler peak in the standard theory) 
is explained in terms of rich clusters of galaxies, 
and also other peaks are predicted.
Angular fluctuations do not depend on frequency, but in the Quasi Steady
State Theory, the relevant opacity (due to carbon whiskers, etc.)
depends strongly on frequency. An a priori prediction of the second
peak is made without dark matter with the MOND (Modified Newtonian
Dynamics\cite{McG99}), and the agreement is good: a very low value
of the amplitude of this peak.
 
Apart from the cosmological hypothesis, the
possible sources for the anisotropies are varied (apart also from the dipolar
component due to the motion of the Earth with respect to the rest of the system):
variations of potential along the background photon path, Sunyaev-Zel'dovich
effect (an inverse Compton effect along the path of the photon), reionization,
photons from other very low temperature sources, Galactic contamination,
intergalactic clouds contamination, etc. At present, all the contaminants are
claimed to be controlled and to be very low and possibly it is so.

In 1999 it was claimed\cite{Lop99b} that Galactic contamination
was not being correctly calculated, due mainly to the fact that templates
of far-infrared dust emission were being used to calculate the fluctuations
of the microwave dust emission  without taking into account the different
temperature of the different regions of the sky (the Galactic dust map in
microwave was being taken as the Galactic dust template in 
infrared multiplied by a
factor depending on the frequency, but independent of the direction).
Taking into account that there are some clouds colder than the average 
temperature as it is really observed\cite{Lop99b} the amplitude of $<TT>$ might 
increase an order of magnitude. Adding rotational dust emission could 
emulate a black body frequency temperature of the anisotropies in the range 
50-90 GHz \cite{Lop99b}, as observed with COBE-DMR.
Apparently, these remarks were unfruitful since all the corrections of the
galactic contamination continue to be carried out in the same wrong way:
using templates and/or constant spectral index of the galactic dust
contamination. Nonetheless, subsequent observations, like those of
Boomerang balloon-borne\cite{deB00}, have proven that, at least in some regions
of low Galactic contamination, the microwave
background radiation anisotropies over a larger range of frequencies is
frequency independent, so something which is not dust
should be producing these anisotropies,
probably the same source which emits the average CMBR. 

\section{NUCLEOSYNTHESIS}
\label{.nucleo}

In the standard model, it is claimed that $^4$He and other light elements 
(Deuterium, $^3$He, $^9$Be, $^7$Li) were created
in the primordial Universe, and the existence of these elements is used as a proof
for the necessity of this hot Universe in the past. However, there are alternatives.
The alternatives may have some caveats to explain accurately all observations,
but neither Big Bang theory is free of problems.

Helium could be created with several explosions such as those in
the (quasi-)steady state theory, or could be synthesized in massive objects evolving 
in the nuclear regions of galaxies\cite{Bur71}.
The existence of massive stars in the first moments of the formation of galaxies, which
in few hundred million years would produce the 24\% helium now observed and
this would be distributed through the interstellar space by supernova
explosions. Indeed, Burbidge \& Hoyle\cite{Bur98} have argued more recently
that a case can now be made for making all the light nuclei in stars.
Population III stars are now believed to contribute to the observed 
Near Infrared Background and heavy element pollution of the intergalactic medium,
and it could contribute to the primordial He abundance\cite{Sal03}. 
Some theoreticians object that there should be more oxygen and carbon
in such a case than observed; this could be solved if the least massive
stars had not exploded but blow off their outer layers (pure helium)\cite{Ler91}(ch. 6).
Certain rare light isotopes cannot have 
been produced in this way, but the cosmic rays generated by early
stars, colliding with the background plasma, would generate them.
Plasma cosmology also produces good fits\cite{Mey03}.
Therefore, there are alternatives to the Big Bang to produce any of these
elements, although, of course, the standard model is the most complete in details
proposal up to now.

$^4$He and $^7$Li abundances are all consistent with those expected a minutes after
the Big Bang, provided that the present universe has baryon density in the
range $0.018<\Omega _bh^2<0.022$ \cite{Sch98}.
The best known abundance is the one for $^4$He, around 0.24 when the metallicity
tends to zero. However, galaxies of poor metal content, 
implying minimal stellar contamination,
show the mass fraction of $^4$He as low as 0.21 \cite{Ter92}.
Even allowing for error bars this is far too low to match the Big Bang 
predictions. Only an ad hoc explanation of inhomogeneities in the
primordial setup conceals the big bang nucleosynthesis with this result.
There are also stars in our own Galaxy which have a low helium abundance in 
their atmosphere: the subdwarf B stars, but it was shown\cite{Sar67} 
that these also showed other peculiarities, phosphorus
and elements like that, which were associated with the chemically peculiar 
stars of population I. It was believed that these objects do not show
a normal Helium abundance and were not taken into account.

The Deuterium evolution and mechanisms of possible Deuterium production
from the Big Bang until now are still not properly understood\cite{Vid98,Pro03}.
For the Beryllium, excessive abundances are observed in the stars\cite{Cas97}, 
but this is justified with the argument that they have no null
metallicity and the abundance should be lower in the primordial Universe.
It is also argued that our Galaxy produces Beryllium, accretion of matter\cite{Cas97},
or that primordial nucleosynthesis 
should be a posteriori substituted by inhomogeneous nucleosynthesis\cite{Boy89}.

Another problem had risen some years ago, but it is claimed nowadays to be solved.
The minimum lower limit to the baryon fraction in clusters of galaxies is around
$\frac{\Omega _b}{\Omega _m}=0.07\pm 0.007$ \cite{Evr96}
so either the density of the Universe
is much lower than $\Omega _m=1$ \cite{Whi93,Whi94,Bag96}, 
or there is an error in the standard 
interpretation of the element abundances or the abundances of the elements
are erroneously measured. The solution nowadays stems from an agreement
that $\Omega _m\approx 0.3 $ and a non-null cosmological constant 
of $\Omega _\Lambda \approx 0.7$ gets that $\Omega =1$ (see \S \ref{.cosmconst}). 

Therefore, we see that the standard model is not free of caveats and many
data are not direct but depend on some other models of elements production and
evolution. Perhaps the most important aspect to criticize is the methodology itself to test the primordial
nucleosynthesis\cite{Kur92}: Instead of trying to make predictions that can
be tested by observation, cosmologists take observations that can be made, such
as measurements of the abundances in Population II material, and try to
determine the primordial abundances, knowing a priori what answer is needed
to fit some particular model. 
Then, they take those derived abundances
and adjust the model to match more closely so that circularity is completely
guaranteed. By this way, the baryon density derived is too low to account
for the subsequent large scale structure of the universe and an ad hoc
addition of cold, dark non-baryonic matter, cosmological constant must be introduced.

There are several free parameters: neutron decay time, number of neutrino species,
ratio between baryons and photons; although perhaps the first two are more or
less known. However, a much better fit to primordial nucleosynthesis
with the observations is given when the number of neutrino species is two instead of three
predicted by the standard model for particles\cite{Hat95}. 

Still, as we see, some open questions are discussed
in the model and whether this can be substituted by alternative models remains
to be seen.

\section{FORMATION OF GALAXIES AND DARK MATTER PROBLEM}

\subsection{Large scale structure}

Like in all the other sections,
there are other theories which explain the formation of voids and
structure: radiation given off by primordial galaxies and QSOs\cite{Kur92};
galaxies which beget other galaxies\cite{Bur97b}; 
plasmas in the middle of magnetic fields and electric currents which create
filaments and this would explain the filamentary Universe\cite{Ler91,Bat97a,Flo97,Bat97b,Bat98};
the Quasi steady State Theory\cite{Nay99}, which reproduces the observed 
two--point correlation function; and others.

Even if we ignore the alternative scenarios, we cannot say that everything
is well undestood in the standard one. Caveats or open questions are still present.
Of course, problems are expected since the total understanding of the phenomenon
is difficult too, and the numerical simulations which are carried out to ascertain
the different characteristics of the large scale structure may not be enough.
As was said by Disney\cite{Dis00} in an joking way: 
{\it ``What about the gas-dynamics, the initial conditions,
the star-formation physics, evolution, dust, biasing, a proper correlation
statistic, the feed back between radiation and matter...? Without a good
stab at all these effects `dotty cosmology' is no more relevant to real 
cosmology than the computer game `Life' is to evolutionary biology''}.
Analytical methods\cite{Bet00,Bet02} to calculate the evolution of the structure
in the non-linear regime have also their limitations.

The standard theory would require large-scale homogeneity on scales of 
distance larger than few tens of Mpc, and the distribution of
galaxies and clusters of galaxies should be random on large scales. However,
the departures from homogeneity claiming a fractal Universe and the regularity
of structure are frequent.
Pencil-beam surveys show large-scale structure out to distances of more 
than 1 Gpc in both of two opposite directions from us. This appears as a 
succession of wall-like galaxy features at fairly regular intervals, with 
a characteristic scale of 128 $h^{-1}$ Mpc\cite{Bro90}, the 
first of which, at about 130 Mpc distance, is called ``The Great Wall''. 
Several such evenly-spaced ``walls'' of galaxies have been found\cite{Kur90}. 
The apparent lack of periodicity in other directions led to the initial
report being regarded as a statistical anomaly\cite{Kai91}, but
a reconfirmation of a $120\pm 15h^{-1}$ Mpc periodicity for clusters
of galaxies came after that (e.g., \cite{Ein97}).

Some observations show large scale inhomogeneities and peculiar velocities typical from a fractal 
on scales up to at least 100 Mpc, so the density fluctuations inside the
fractal inhomogeneity cell will lead to strong disturbance of pure Friedmann
behaviour\cite{Hag72,Fan91,Rib92a,Rib92b,Rib93}. 
Therefore, the Universe is not even close to the 
Friedman--Lema\^\i tre--Robertson--Walker on scales less than 100 Mpc.
In Las Campanas redshift survey, statistical differences from homogeneous 
distribution were found out to a scale of at least 200 Mpc\cite{Bes00}. 
Structure is dominated by filaments
and voids on this level\cite{deL86} and by large velocity
flows relative to the cosmological background\cite{Dre87}.
However, other observations suggest the opposite conclusion:
a striking linearity of the Hubble law in the distance range between
2 and 25 Mpc\cite{San86,Pee98}. This is the inhomogeneity paradox\cite{Bar94}. 
Beside the paradox, parameters like $H_0$, $q_0$
and $\Omega _0$ are determined without knowing in which scales the radial
motion of galaxies and clusters of galaxies relative to us is completely
dominated by the Hubble flow\cite{Mat95}.
Most cosmologists appeal to the highly isotropic character of
the microwave background as one of the principal justifications for
assuming that the Universe is homogeneous on large scales. By itself,
the fact that some observer sees isotropic background radiation is
inconclusive: for this can be true in a static inhomogeneous universe, as well
as in a spherically symmetric inhomogeneous universe where we are near
a centre of symmetry\cite{Mat95}. So an open question
is apparently present.

The local streaming motions of galaxies are too high for a finite universe 
that is supposed to be everywhere uniform.
In the early 1990s, we learned that the average redshift for galaxies of a 
given brightness differs on opposite sides of the sky\cite{Mat92,Lin92,Fin93,Hud99}. 
There is streaming inconsistent with the Local Group absolute space velocity 
inferred from the CMB dipole anisotropy\cite{Lau94}.
The standard model interprets this as the existence of a puzzling group 
flow of galaxies relative to the microwave radiation on scales of at least 
130 Mpc. Earlier, the existence of this flow led to the hypothesis of a 
``Great Attractor'' pulling all these galaxies in its direction. But in newer 
studies, no backside infall was found on the other side of the hypothetical 
feature. Then, the only alternative within the standard model to the apparent 
result of large-scale streaming of galaxies 
is that the microwave radiation is in motion relative to us. Either way, this 
result is trouble for the orthodox interpretation.

In a homogeneous and isotropic universe we also expect the redshift distribution
of extragalactic objects to approximate to a continuous and
aperiodic distribution. However,
Tifft\cite{Tif74,Tif76,Tif77,Tif80}  affirmed that there was a periodicity of 
70-75 Km/s in the redshift
of the galaxies. A periodicity with $\Delta z=0.031$ or 0.062
is also found for the QSOs\cite{Bur72,Bel02b}. The non-uniformity,
peaking or periodicity in the redshift distribution of QSOs and the Tifft-effect are at 
present two unrelated phenomena. However, none of them can be understood
in terms of the Cosmological Hypothesis.
In an improved correction
for the optimum solar vector, the periodicity is found to be 37.22 Km/s, with a
probability of finding this period by chance of $2.7\times 10^{-5}$ \cite{Gut96}.

\subsection{Antimatter}

Protons do not decay, as far as we know from experiments of particle physics
until now. Therefore,
the Universe should be made up equally of matter and antimatter according to
the standard model. Matter dominates the present universe apparently 
because of some form of asymmetry, such as CP violation asymmetry, that caused 
most anti-matter to annihilate with matter, but left much matter. Experiments 
are searching for evidence of this asymmetry, so far without success. Other 
galaxies cannot be antimatter because that would create a matter-antimatter 
boundary with the intergalactic medium that would create gamma rays, which 
are not seen\cite{Tau97}. Another open question.

\subsection{Dark matter}

Stellar and cold gas in galaxies sum a baryonic matter 8$^{+4}_{-5}$\%
of the total amount of the Big Bang baryonic matter predictions\cite{Bel03}, 
where is the rest of the baryonic material? From here
stems another open question. And the baryonic matter is only around a tenth part
of the total amount of matter\cite{Tur02}. 
Which is the nature of the putative non-baryonic dark matter
to achieve the current value of $\Omega _m\approx 0.3$?

Current CDM models predict the existence 
of dark matter haloes for each
galaxy whose density profiles fall approximately as $r^{-2}$, although the
original ideas\cite{Whi78} concerning hierarchical structures with
CDM, which gave birth to the present models, was that the dark matter was
distributed without internal substructure, more like a halo with galaxies
than galaxies with a halo\cite{Bat00}, something similar to the scenario
in \cite{Lop99,Lop02b}. 
Some authors have been led to question the very existence 
of this dark matter on galactic scales since its evidence is
weak\cite{Bat00,McG00,Eva01,Tas02} and the predictions do not fit the
observations: the halos are not cusped as predicted,
the predicted angular momentum is much less than the observed one, 
the number of satellites around the galaxies is much less than the predictions,
etc. Microlensing surveys\cite{Las00} constrain the mass of the halo in our Galaxy
in the form of dim stars and brown dwarfs to be much less than that
necessary for dark matter haloes.
Some observations are inconsistent with the dominant dark matter component
being collisionless\cite{Moo94}. Neither are black hole haloes  a consistent
scenario\cite{Moo93}. The nature of the dark matter 
has been investigated\cite{Sad99} and there are no suitable candidates.
Moreover, some other dynamical problems can be solved without dark matter:
galactic stability\cite{Too81}, warp creation\cite{Lop02b}, rotation 
curves\cite{Bat00,San02}. Velocities in galaxy pairs or satellites 
might measure the mass of the intergalactic 
medium filling the space between the members of the 
pairs\cite{Lop99,Lop02b} rather than the mass of dark halos
associated to the galaxies. 

Dynamical mass in clusters is also used to claim the existence of the 
dark matter, and it is likely that much of the dark matter be intercluster matter, in rich clusters
or also some amounts in small clusters like the Local Group\cite{Lop99}.
The virial theorem applied to clusters of galaxies, 
which implies there must be a large amount of dark matter, 
perhaps a 90\% of the mass of the cluster, might be not applicable
since the virial theorem is suspect to be not applicable to these
clusters\cite{Kur92}. However, X-ray observations and gravitational lensing
give masses not very far from those derived with the virial theorem, so
this must not be very wrong. Also,
in many numerical simulations dwarf galaxies are less strongly clustered
than giants\cite{Pee98}: there is a biasing; the galaxies are fair
tracers of mass for the purpose of estimating $\Omega _m$, so dynamical
studies may be not appropriate to get $\Omega _m$ from the clusters observations.
In any case, it seems clear that there may be some matter which we do not see.

There is some dark matter, it is clear, but how much, and what is its nature?
The success of the standard model to convert a hypothesis into a solid theory
depends strongly on the answer to these open questions.

\subsection{Cosmological constant}
\label{.cosmconst}

The question of the cosmological constant\cite{Pad03}, Einstein's biggest blunder considered
now to be not such a blunder, is very funny.
Few years ago, most cosmologist thought the scenarios dominated by
the cosmological constant were not favoured\cite{Fuk91}.
In the eighties, the cosmological constant was many times disregarded as an 
unnecessary encumbrance or its value was put to zero\cite{Lon86},
and all the observations gave a null or almost null value.
However, since other problems in cosmology have risen, many cosmologists
realized that a $\Omega _\Lambda =0.70-0.80$ could
solve many of them in CDM cosmology\cite{Efs90} at
the beginning of the 90s. Years later, the evidences for such
a value of the cosmological constant began to come. A genial prediction
or a prejudice which conditions the actual measures? Another open question.

One of the measures of the cosmological constant comes nowadays from
the supernovae, which give some defect of luminosity in distant
sources which can be justified with the introduction of the cosmological
constant. It was criticized to be due possibly to
intergalactic dust\cite{Agu00,Goo02}. The presence of grey dust is
not necessarily inconsistent with the measure of a supernova at z=1.7
(SN 1997ff)\cite{Goo02}.
Type Ia Supernovae have also possibly a metallicity dependence and
this would imply that the evidence for a non-zero cosmological constant 
from the SNIa Hubble Diagram may be subject to corrections for metallicity
which are as big as the effects of cosmology\cite{Sha01}. 
Also, there was an underestimate of the effects of host galaxy extinction: a 
factor which may contribute to apparent faintness of 
high-z supernovae is evolution of the host galaxy extinction with z \cite{Row02}; 
therefore, with a consistent treatment of host galaxy
extinction and elimination of supernovae not observed before maximum, 
the evidence for a positive $\Lambda $ is not very significant.

There are other sources of $\Omega _\Lambda $ measure: the anisotropies
of the microwave background ra\-dia\-tion\cite{Ben03,Rub03} 
(see \S \ref{.anis}), for instance.
In the last 6-7 years, a lot of proofs have been presented to the community
to convince us that the definitive cosmology has $\Omega _\Lambda \approx 0.7$,
which is surprising taking into account that in the rest of the history
of the observational cosmology proofs have been presented for $\Omega _\Lambda \approx
0$. Still some recent tests point out that other values are available in the
literature. For instance, from the test angular size vs. redshift for
ultracompact radio sources, it is obtained that $\Lambda $ has the opposite sign\cite{Jac97}.

Beside the observational facts, other alternative theories can predict 
the measured value. For instance, it was argued\cite{Ban00,Nar02} that
the Quasi Steady State Cosmology provides also a good fit to the
data which support $\Omega _\Lambda =0.7$ due to 
a repulsive force arising from a negative energy C-field,
and a cosmic dust in the form of metallic whiskers which extinguishes radiation
traveling over long distances.

Moreover, the actual values of $\Omega _\Lambda $ have some consistency problem 
in the standard scenario of inflationary Big Bang. 
The cosmological constant predicted by the quantum field theory is
a value much larger than those derived from observational cosmology.
This is due to the fact that the vacuum energy in the quantum field
theory takes the form of the cosmological constant in Einstein's equations.
If inflation took place at the GUT
epoch, the present value is too low by a factor $\sim 10^{-108}$, and
if the inflation took place at the quantum gravity epoch, the above factor is 
lower still at $\sim 10^{-120}$ \cite{Wei89}.

Again, everything is far to be properly understood and with good 
constrained parameters.

\subsection{Distant galaxies and evolution}
\label{.distgal}

The Big Bang requires that stars, QSOs and galaxies in the early universe 
be ``primitive'', meaning mostly metal-free, because it requires many generations 
of supernovae to build up metal content in stars. But the latest evidence 
shows the existence of even higher than solar metallicities 
in the ``earliest'' QSOs and galaxies\cite{Fan01b,Bec01,Con02}. 
The iron to magnesium ratio increases at higher redshifts\cite{Iwa02}.
And what is even more amazing: there is no evolution of some line ratios, including iron 
abundance\cite{Die03,Fre03,Mai03,Bar03}, between $z=0$ and $z=6.5$ .
The amount of dust in high redshift galaxies and QSOs is also much higher than expected\cite{Dun03}.
In view of these evidences, orthodox cosmologists claim now that the star formation began very 
early and produced metals up to the solar abundance quickly, in roughly half Gyr. However, 
it is not enough with such a surprising claim, it needs 
to be demonstrated, and I do not see any evidence in favour of such a quick evolution in the local
galaxies.

Observation does not always match the predictions
neither for the positions of first hydrogen clouds nor for the galaxies.
Examples are the possible discovery of a giant
pancake of hydrogen gas in the early Universe\cite{Uso91}, 
an object that should not exist according to CDM model; and that 
the most distant galaxies in the Hubble Deep Field show insufficient evidence 
of evolution, with some of them having very high redshifts ($z=6-7$).
The presence of diffuse neutral hydrogen should produce an
absorbing trough shortward of a QSO's Lyman-alpha emission 
line---Gunn-Peterson effect.
A hydrogen Gunn-Peterson trough was predicted to be present 
at a redshift $z\approx 6.1$ \cite{Mir00,Bec01}. Indeed,
a complete Gunn-Peterson trough at $z=6.28$\cite{Bec01} was discovered, 
which means that the Universe is approaching the reionization
epoch at $z_r\approx 6$\cite{Bec01}. However, galaxies  
have been observed at $z=6.68$\cite{Lan98}, or $z=6.56$\cite{Hu02} 
without the opacity features prior to
the reionization, and the epoch of reionization was moved beyond $z_r=6.6$\cite{Hu02}. 
An inhomogeneous reionization\cite{Bec01} is a possibility to explain
the apparent disagreement of the different data.
Recent measures of CMBR anisotropies by the WMAP observations give a reionization
epoch $z_r=20^{+10}_{-9}$ (95\% CL)\cite{Ben03}. If we were going to believe that CMBR 
anisotropies are being correctly interpreted in terms of the standard cosmology, 
we would have again a new inconsistency.

With regard the topic of the evolution itself, I will not discuss it
in this review since there are up to date many different claims in the literature
which have not found a clear common position (e.g., \cite{Bec99}). 
Today somebody demonstrates the evolution, tomorrow somebody demonstrates 
there is no evolution. That is the scenario at present. 
Of course, the study of the evolution has intrinsic difficulties
which avoid to extract definitive conclusions: they depend on the assumptions for
the cosmological model, the intergalactic absorption, etc. It is also a difficult
area to extract definitive conclusions.

\subsection{The age of the Universe}

Controversy also surrounds the age of the Universe. Cosmologists say the
Universe is 8 to 15 billion years old, 13.7 Gyr according to the recent measures
of WMAP\cite{Ben03}, while stellar astronomers claim
that the oldest stars in the Galaxy are as high 16 to 19 billion 
years old\cite{Jay93} or even older. 
In any case, I consider that it is still a good argument in favour of the Big Bang 
that the order of magnitude of these ages be roughly equal given the inaccuracies 
of the measures. At present we have no firm proof for the age of 
the stars larger than the predicted age of the Universe by the standard model.

\section{CONCLUSIONS}

I have no conclusions to claim. I think that anybody who reads this paper
can have his/her own conclusions. I just want to quote some sentences that
I liked by Pecker:

\begin{quotation}
{\it ``And we would pretend to understand everything about cosmology,
which concerns the whole Universe? We are not even ready to start to
do that. All that we can do is to enter in the field of speculations.
So far as I am concerned, I would not comment myself on any
cosmological theory, on the so-called `standard theory' less on many
others. Actually, I would like to leave the door wide open.''}\cite{Pec97}
\end{quotation}

I agree: the door wide open. Some important caveats are still present in the
standard scenario, and even if many of the observations reviewed here are not correct,
there remain still many others which are. There are still many open questions
to consider that we have in our hands a final theory with only some parameters
which remains to be fitted. Possibly the own foundations of the standard theory are wrong, or
possibly they are in the good way. Of course, criticizing is easier than building a
theory, and the achievements of the standard theory must not be 
underestimated, but I think it is too early to close the doors behind us.

\

{\bf Acknowledgements:}
Thanks are given to Carlos M. Guti\'errez (Inst. Astrof. Canarias, 
Tenerife, Spain), Niranjan Sambhus (Astr. Inst. Basel, Switzerland), Juan
E. Betancort-Rijo (Inst. Astrof. Canarias) and
Halton C. Arp (Max Planck Inst., Garching, Germany)
who read the manuscript and gave helpful comments.
I also thank the proof-reading of this paper from Anna Mar\'\i a D'Amore 
(cetet@ciu.reduaz.mx, CETET, M\'exico) and Niranjan Sambhus.

\end{document}